\documentclass[aps,prl,twocolumn,showpacs,preprintnumbers,amsmath,amssymb]{revtex4-1}
\usepackage{graphicx}
\usepackage{color}
\usepackage{dcolumn}
\usepackage{bm}

\begin{document}

\preprint{}

\title{Dark matter in the hidden gauge theory}

\author{Nodoka~Yamanaka$^1$}
  \email{nodoka.yamanaka@riken.jp}
\author{Sho~Fujibayashi$^2$}
\author{Shinya~Gongyo$^{2,3}$}
\author{Hideaki~Iida$^2$}
  \affiliation{$^1$iTHES Research Group, RIKEN, 
  Wako, Saitama 351-0198, Japan}
  \affiliation{$^2$Department of Physics, Graduate School of Science,
  Kyoto University, 
  Kitashirakawa-oiwake, Sakyo, Kyoto 606-8502, Japan}
  \affiliation{$^3$Department of Physics, New York University, New York, 10003, USA}

\date{\today}

\begin{abstract}
The cosmological scenario of the dark matter generated in the hidden gauge theory based on the grand unification is discussed.
It is found that the stability of the dark matter halo of our Galaxy and the cosmic ray observation constrain, respectively, the dark matter mass and the unification scale between the standard model and the hidden gauge theory sectors.
To obtain a phenomenologically consistent thermal evolution, the entropy of the standard model sector needs to be increased.
We therefore propose a scenario where the mini-inflation is induced from the potential coupled to the Standard model sector, in particular the Higgs sector.
This scenario makes consistent the current dark matter density as well as the baryon-to-photon ratio for the case of pion dark matter.
For the glueball or heavy pion of hidden gauge theory, an additional mini-inflation in the standard model sector before the leptogenesis is required.
We also propose the possibility to confirm this scenario by known prospective experimental approaches.

\end{abstract}


\maketitle


From recent observations, it is now known that 27\% of the energy of our Universe is composed of dark matter (DM) \cite{planck}.
The presence of it was also indicated by many previous observations \cite{zwicky,davis,clowe}, and the result of the N-body simulation suggests that this medium forms nonrelativistic clusters \cite{navarro}, which explains the DM halo surrounding our Galaxy.
We currently believe that this halo is indispensable for the formation of stars and galaxies that we can observe today \cite{structureformation}, and consequently for the origin of the life and our existence.
The investigation of the origin of the DM is thus one of the most essential subject.

Currently, the DM is known to be composed of weakly interacting massive particles \cite{macho}.
However, it is also currently known that there are no candidates of DM particles in the standard model (SM) of particle physics.
We therefore need to introduce a new extended theory beyond the SM to explain the DM \cite{dmreview,susydm,inertdm,extradm,axiondm,sterileneutrinodm}.
Here we would like to investigate a natural scenario for the inclusion of the DM by postulating the existence of new gauge forces which are unified at the fundamental scale but decoupled with the visible sector.
This {\it hidden gauge theory} (HGT) presents many strong advantages in the point-of-view of the naturalness: 1) The fundamental interactions are unified at some high energy scale, like the commonly believed grand unified theory (GUT) of the SM sector \cite{gut}, and this fact allows the existence of many gauge interactions out of the SM sector. Moreover, this high energy scale guaranties the weak correlation with the SM sector.
2) The mass scales of the gauge theory are controlled by the color and flavor numbers through the running coupling, so that the theory does not involve a serious hierarchical problem.
3) The lightest particles are DM hadrons, so they can be natural candidates of DM component, although being self-interacting \cite{carlson,secluded,hidden,compositedm,chiral,glueballino}.

In this letter, we discuss the DM generated by the additional hidden $SU(N_c)$ gauge theory which is  unified with the SM sector at some GUT scale.
We first examine the cosmological dynamics of the HGT and the phenomenological constraints on it.
We then propose a new scenario requiring mini-inflations to consistently implement the HGT into the cosmology.
The prospective approaches to confirm this scenario are finally given.
In this work, we do not discuss the mechanism for generating the gauge theories at the GUT scale.
We only consider $SU(N_c)$ gauge theories with $N_f$ fermions, without any scalar bosons and supersymmetry \cite{glueballino}.


In this work, we assume the $SU(N_c)$ HGT with $N_f$ equal mass fermions.
This nonabelian gauge theory dynamically generates a mass scale $\Lambda_{\rm DM}$.
In the HGT, the lightest particle is a pion if there is at least one quark lighter than the scale parameter $\Lambda_{\rm DM}$, or a glueball if there are no quarks lighter than $\Lambda_{\rm DM}$ \cite{chiral}.
From a simple dimensional analysis, the mass of the DM glueballs, $m_\phi$, and that of the DM pions, $m_\pi$, in the $SU(N_c)$ gauge theory are respectively given by
\begin{eqnarray}
m_\phi \sim \Lambda_{\rm DM} , \ \ m_\pi
=
\frac{1}{f_{\rm DM}} \sqrt { m_q \langle 0 | \bar q q |0 \rangle }
\sim
\sqrt{ m_q \Lambda_{\rm DM}}
 ,
 \ \ \ \ 
\label{eq:pionmass}
\end{eqnarray}
where $f_{\rm DM}$ and $\langle 0 | \bar q q |0 \rangle$ are, respectively, the pion decay constant and the chiral condensate, which is approximated to be equal to $ \Lambda_{\rm DM}$ and $ \Lambda_{\rm DM}^3$.

\begin{figure}[htb]
\includegraphics[width=5cm]{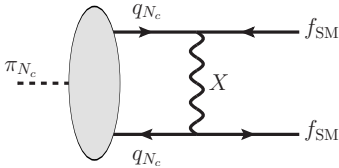}
\caption{\label{fig:DM_decay}
Examples of diagram contributing to the decay of DM to SM particles.
The SM fermion is denoted by $f_{\rm SM}$, the hidden quark by $q_{N_c}$, and the GUT gauge boson by $X$.
}
\end{figure}

We first see the constraint on the scale of unification between the SM and the HGT.
As the HGT and the SM are unified at $\Lambda_{\rm GUT}$, DM particles interact with SM particles through gauge bosons with mass of $O( \Lambda_{\rm GUT})$.
The clearest way to find the unification scale is to directly test the production of DM via the accelerator \cite{lhc} or direct DM detection experiments \cite{directdetection}.
These set a constraint of $\Lambda_{\rm GUT} > O ({\rm TeV})$ for $m_{\rm DM} < O({\rm TeV})$.
The most sensitive approach to $\Lambda_{\rm GUT}$ is the indirect detection experiments \cite{ams-02,pamela,fermilat,icecube}, where the decay product of DM particles are observed as cosmic rays (CR).
A typical process is given in Fig. \ref{fig:DM_decay}.
In the HGT scenario, the indirect search is much more sensitive than the direct detection, since the SM particles are produced via decays.

The detection rate (flux) of the neutral decay products can be estimated by dimensional analysis, as
\begin{equation}
\frac{d \phi}{d E}
\sim \frac{1}{4\pi} \frac{m_{\rm DM}^3 }{m_X^4} \rho_0 R\, \delta (E-m_{\rm DM}/2)
,
\end{equation}
where $m_{\rm DM}$ is the mass of the lightest DM hadron, and $R \sim 20$ kpc the radius of the halo.
Here we have assumed a flat distribution of the DM halo with the density $\rho_0 \sim 0.3\, {\rm GeV}/{\rm cm}^3$, the local DM density near the Earth.
The density should be larger near the center of the galaxy \cite{navarro}, so this estimation is conservative.
In this work we have only considered high energy neutrinos.

\begin{figure}[htb]
\includegraphics[width=8cm]{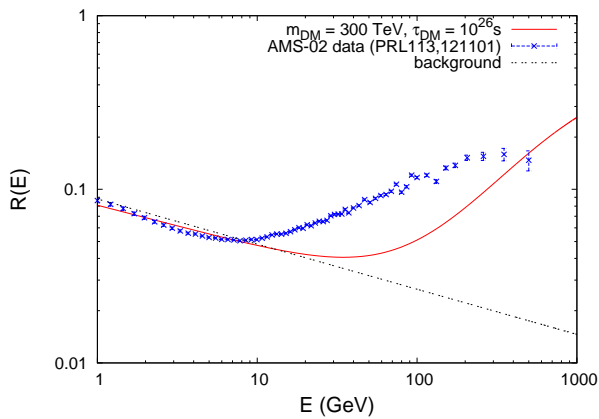}
\caption{\label{fig:positron_fraction}
Positron fraction ($R \equiv \frac{N_{e^+}}{N_{e^-} +N_{e^+}}$) of the CR.
}
\end{figure}

For charged CR, the diffusion through galactic diffusion zone must be considered \cite{strong,baltz,ibe}.
In this work, we only consider electron/positron pair production via 2-body decay.
We have calculated the spectrum of the positron from DM decay following Ref. \cite{baltz}.
The result is shown in Fig. \ref{fig:positron_fraction}.
We have used the background contribution of Ref. \cite{ibe}.

The origin of the high energy CRs is not yet determined.
In this work, we just set the constraint on the unification scale by imposing that the flux of the decay products should not exceed the observed data.
From the experimental data of electron/positron detection of AMS-02 \cite{ams-02} and those of high energy neutrino detection of IceCube \cite{icecube}, we obtain roughly the constraint
\begin{equation}
\Lambda_{\rm GUT} > O(10^{14}) {\rm GeV}
,
\end{equation}
for $m_{\rm DM} < 1$ PeV.
This result thus suggests that the GUT scale is located at a very high energy scale.


We now constrain the DM mass from the thermal history.
The HGT phase transition occurs at $T_c \sim \Lambda_{\rm DM}$, but it is not important in our work since there the shift of the degrees of freedom is $O(1)$.
Rather the nonrelativistic transition at $T \sim m_{\rm DM}$ has important effect on the thermal evolution of the Universe, since the relativistic degrees of freedom controls the expansion rate.
We further neglect all changes of degrees of freedom due to the nonrelativistic transition.

When there is no DM baryon number asymmetry, the relic density of the DM hadrons is approximately
given by
\begin{eqnarray}
n_{\rm DM} (T = m_{\rm DM} )
&\sim &
O(m_{\rm DM}^3 )
.
\label{eq:boltzmann}
\end{eqnarray}
In this relation, there is no suppression from the Boltzmann factor, because the HGT is not correlated with the other sectors, and no transfer of entropy discussed in Refs. \cite{secluded} occurs.
At $T < m_{\rm DM}$, the energy density decreases like nonrelativistic matter even if the particles are self-interacting, as the correction of the scale factor dependence is only logarithmic \cite{carlson}.
If Sakharov's criteria \cite{sakharov} are fulfilled, the HGT baryons remain as thermal relic.
In this case, interesting dark nuclear phenomena may occur \cite{mccullough}.
However, as Sakharov's criteria are controlled by other inputs, we will not discuss the baryon relic scenario further.

From the current DM energy fraction, Eq. (\ref{eq:boltzmann}) and 
the entropy conservation ($T^3 a^3 \approx T_{\rm eq}^3 a_{\rm eq}^3$,
where $a$ is the scale factor of the Universe, and the subscript ``eq'' denotes the physical quantities at the time of matter-radiation equality),
we can give the DM particle mass as
\begin{equation}
m_{\rm DM}
= 
\xi \times
O(10^{-8}) {\rm GeV}
,
\label{eq:naivetfo}
\end{equation}
where the parameter $\xi$ is the entropy ratio between the SM sector and the DM sector.
If the entropies of the SM and DM sectors are not disturbed during their evolution, $\xi \sim 1$,
and the DM freeze-out occurs near the temperature of recombination.
This upsets the structure formation (hot DM problem) due to the erasure of fluctuations, and also the big bang nucleosynthesis, which is quite sensitive to the expansion rate near $T \sim O({\rm MeV})$.
We will see next that the strongest constraint is given by the stability of the halo.

\begin{figure}[htb]
\includegraphics[width=7cm]{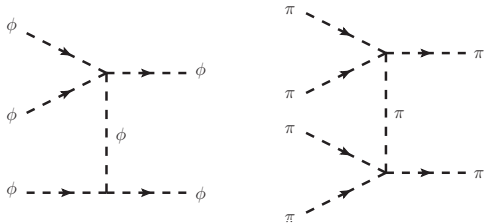}
\caption{\label{fig:annihilation}
Examples of diagrams contributing to the number reducing processes.
The DM glueball is denoted by $\phi$, and the DM pion by $\pi$.
}
\end{figure}

The current DM particles are bound in the gravitational potential of the galactic halo with a particle density much larger than the mean density of the Universe, but the halo is stable at least at the scale of the age of the Universe.
This fact provides a strong constraint on the parameter of the HGT, since the DM particles can reduce their number through the many-body annihilation, shown in Fig. \ref{fig:annihilation}, and relativistic expulsion from the halo.
These processes contribute to the {\it decay of the halo}, and the decay rate should not be larger than the current Hubble constant.

To estimate the halo decay rate we assume the following low energy effective lagrangian of the HGT \cite{glueballino,chiral}
\begin{align}
{\cal L}_\phi 
&=
\partial^\mu \phi^\dagger \partial_\mu \phi 
- \frac{m_\phi^2}{2} \phi^2
- \frac{A_\phi}{3!} \phi^3
- \frac{\lambda_\phi}{4!} \phi^4
,
\\
{\cal L}_\pi 
&=
\frac{f_{\rm DM}^2}{4}{\rm Tr} \Bigl[ \partial^\mu U^\dagger \partial_\mu U \Bigr]
+ \frac{f_{\rm DM}^2 B }{2} {\rm Tr} \Bigl[ M U^\dagger + U M^\dagger \Bigr]
,
\end{align}
where $M$ the pion mass matrix in the $SU(N_f)$ chiral effective theory.
The parameter $B$ is the linear coefficient of the current quark mass.
In this work, we assume that $M_{ab} = m_\pi \delta_{ab}$.

The leading processes contributing to the decay rate, the glueball $3\rightarrow 2$ annihilation in the case without light fermions, and the pion $4\rightarrow 2$ annihilation where $m_q < \Lambda_{\rm DM}$,
can be estimated by order analysis as
\begin{eqnarray}
\Gamma_3
&\sim &
C_\phi \frac{A_\phi^2 \lambda_\phi^2}{m_\phi^{10}} \rho_0^3 R^3
\sim
C_\phi \frac{\rho_0^3 R^3}{\Lambda_{\rm DM}^8}
,
\label{eq:Gamma_3}
\\
\Gamma_4
&\sim &
C_\pi \frac{ \rho_0^4 R^3}{f_{\rm DM}^8 m_\pi^4}
\sim
C'_\pi \frac{\rho_0^4 R^3}{m_q^2 \Lambda_{\rm DM}^{10}}
.
\label{eq:Gamma_4}
\end{eqnarray}
The constants $C_\phi$ and $C_\pi$ ($C'_\pi$) represent the enhancement factors.
From large $N_c$ and $N_f$ analysis, $C'_\pi \propto N_c^{-4} N_f^{-2}$. 
We see that the annihilation process is attenuated when the flavor number increases.
In the present case, the Sommerfeld effect \cite{sommerfeld,arkani-hamed} is not important, since the interaction range and the hadron mass are close.

\begin{figure}[t]
\includegraphics[width=7.5cm]{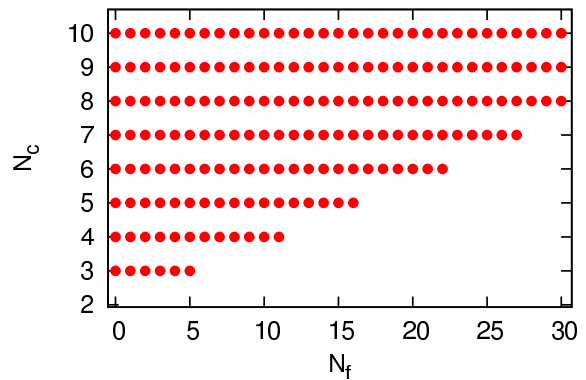}
\vspace{-0.4cm}
\caption{\label{fig:NcNf}
Allowed $N_c$ and $N_f$ of HGT (red points) with $\Lambda_{\rm GUT}=10^{16}{\rm GeV}$ and $m_q=0$, constrained by Eqs. (\ref{eq:Gamma_3<H_0}) and (\ref{eq:Gamma_4<H_0}).
}
\end{figure}

To obtain a stable DM halo, we must have $\Gamma < H_0$ where $H_0 
\sim 70 \,{\rm km\, s}^{-1}\, {\rm Mpc}^{-1}$ is the current Hubble expansion rate.
We then obtain
\begin{eqnarray}
\Gamma_3 < H_0
&\Rightarrow &
\Lambda_{\rm DM}
>
10^4 \, {\rm GeV},
\label{eq:Gamma_3<H_0}
\\
\Gamma_4 < H_0
&\Rightarrow &
\Lambda_{\rm DM}
>
\Bigl( \frac{\Lambda_{\rm DM}}{m_q} \Bigr)^{\frac{1}{6}} \times 10^{-1} \, {\rm GeV}
.
\label{eq:Gamma_4<H_0}
\end{eqnarray}
The first inequality corresponds to the case where the glueball is the lightest DM, and the second one to that for the pion.
Note 
that increasing the flavor number will sizably loosen the constraint (\ref{eq:Gamma_4<H_0}).
By comparing with the thermal extrapolation result (\ref{eq:naivetfo}), we see that the entropy of the SM sector need to be enhanced at least by $\xi \sim 10^{12}$ for DM glueball, and by $\xi \sim 10^7$ for DM pion (the current fermion mass dependence is small), compared with the DM sector entropy.
Eqs. (\ref{eq:Gamma_3<H_0}) and (\ref{eq:Gamma_4<H_0}) can also constrain $N_c$ and $N_f$ combined with the perturbative running coupling analysis (see Fig. \ref{fig:NcNf}).
We see that $N_c = 2$ hadrons as the main component of DM is excluded.

Here we point that if some flavor numbers are nonzero with vanishing total flavor number, Bose-Einstein condensation of DM pions occurs in the halo \cite{bec}.
In that case, the NG modes (pions) reduce their number \cite{ngmodes}.
The decay rate of the halo may therefore be enhanced, and the phenomenological constraint on the mass scale of the HGT is strengthen.
This interesting situation has chances to be realized when there are C and CP violations in the HGT.


We now discuss how to enhance the entropy of the SM sector, and to fulfill Eqs. (\ref{eq:Gamma_3<H_0}) and (\ref{eq:Gamma_4<H_0}).
The most ad hoc way to solve this conflict is to impose an asymmetry between the entropy of the DM and SM sectors, for instance via asymmetric decay of inflatons.
This scenario however needs additional fine-tuning in the models, since the entropies of these two sectors differ by more than $O(10^{7})$, and is thus difficult to keep the naturalness. 

The natural resolution we propose here is a scenario where some mini-inflations \cite{inflation}
are induced from potentials coupled to the SM sector.
We note that the multi-inflationary scenario is totally natural, since we are encountering now the accelerating expansion \cite{darkenergy}.
Although the SM Higgs potential with the Higgs boson mass $m_H = 126$ GeV \cite{higgs} cannot generate inflation, the SM Higgs sector has a serious hierarchical problem and the necessity to extend it may allow a mini-inflation.
If a mini-inflation occurs at the Higgs transition, the DM density will be diluted, and the entropy of the SM sector will be enhanced due to the reheating at the end of the inflation.
The mini-inflation is of course not restricted to the Higgs scale, and other candidates such as the QCD phase transition with finite baryon number density \cite{qcdinflation}, or higher scale phenomena, for instance, along the leptogenesis, are possible.

The above scenario has another advantage, because it can also dilute the baryon number asymmetry.
One promising mechanism of baryogenesis is the leptogenesis with Majorana neutrino \cite{leptogenesis}.
There the lepton number asymmetry was created at a very high temperature, of the order of $T \sim \frac{v_H^2}{m_\nu} \sim 10^{12}$ GeV, and the lepton number transferred to the baryon number asymmetry through the sphaleron process \cite{sphaleron}, where $v_H$ is the Higgs vacuum expectation value, and $m_\nu$ the physical neutrino mass.
The natural order of the lepton number asymmetry 
should be $O(1\mathchar`-0.01)$ at the moment of its generation.
Currently, the observed baryon number asymmetry, the baryon-to-photon ratio, is $\frac{n_B}{n_\gamma} = 6 \times 10^{-10} $, and a large amount of dilution is needed.
The mini-inflation of the Higgs sector with the relative expansion of the order $\Delta a  \sim 10^2\mathchar`-10^3$, which brings an enhancement of the entropy of order of $\xi \sim 10^6 \mathchar`-10^9$ will explain the current baryon-to-photon ratio and the DM pion with $m_{\rm DM} \sim$ GeV.
The mini-inflation of the Higgs sector is thus a good candidate for this suppression.

\begin{figure*}[htb]
\includegraphics[width=18cm]{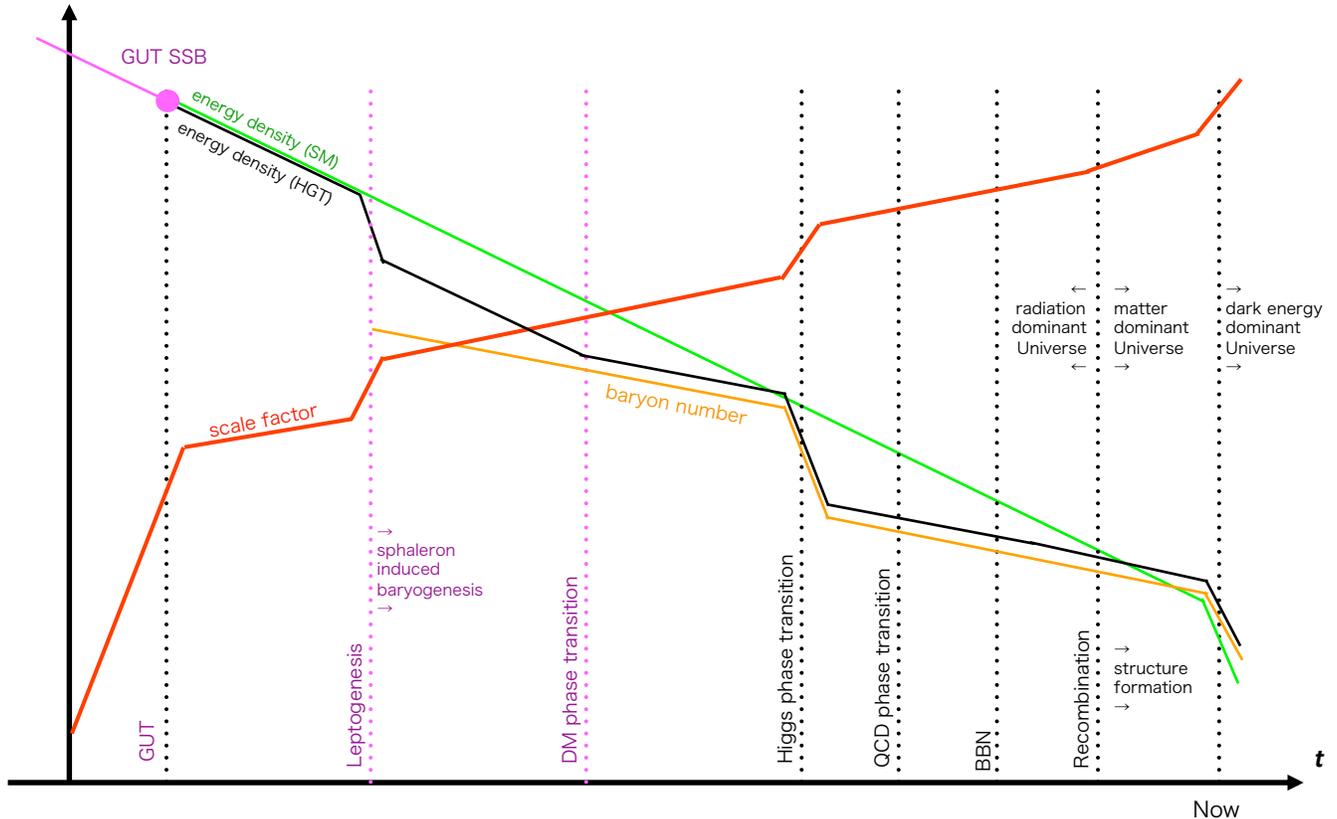}
\caption{\label{fig:thermal_history}
Schematic picture of the thermal history.
}
\end{figure*}

To fulfill the thermal relic density of the DM glueball or DM pion with $m_{\rm DM} > 100$ GeV, a larger mini-inflation is required.
But if the Higgs sector increases the entropy by $\xi > 10^{12}$, the baryon number asymmetry will be more diluted than the observed data.
An alternative resolution is therefore to generate an additional mini-inflation of the order of $\Delta a > 10^{1}$ between the GUT scale and the leptogenesis from the SM sector.
This does not forbid the case where the mini-inflation occurs through a process correlated with the leptogenesis.
An example of the thermal history is shown 
in Fig. \ref{fig:thermal_history}.

We now list the observables sensitive to the scenario discussed and the possibility to determine it in the future.
We have to pay attention on the dynamics of both the HGT and SM sectors.

The direct sensitive probes to the mini-inflation are the inflationary B-modes \cite{bicep}.
For the case with DM glueball or DM pion heavier than $100$ GeV, the mini-inflation has occurred at very high temperature in the range of $ 10^{12}\, {\rm GeV} < T < 10^{15}$ GeV.
The B-modes with signals smaller than the GUT scale inflation by several orders of magnitude is therefore expected.
The B-modes due to the Higgs or QCD scale mini-inflations will be difficult to observe as the energy density of the gravitational waves generated there is very small.
We must also point that if there are more than two inflations in the past, the last inflation will dilute the gravitational waves of other past inflations.
It will therefore be very difficult to observe the B-modes of any origin if the Higgs or QCD scale mini-inflation has occurred.

The second observable to point is the stochastic gravitational wave background radiated through the first order phase transition \cite{grojean,kikuta,envelope}.
During the phase transition, the bubbles collide with each other, and gravitational waves are radiated.
This observable may 
determine the critical temperature and the first order nature of the hadron phase transition of the HGT sector.
In Refs. \cite{grojean,kikuta,envelope}, the explicit analytic formulae for the gravitational wave background were given.
For the pure Yang-Mills HGT, the transition is of first order, so it is possible to probe it by observation.
By a dimensional analysis, the imprint frequency and energy component of the stochastic gravitational waves are given by
\begin{eqnarray}
f\simeq \frac{T_c^2 }{m_{\rm Pl} } \cdot \frac{a (T_c)}{ \Pi_i \Delta a_i } 
,
\ \ \ \Omega_{\rm GW} h^2 \equiv
\frac{\rho_{\rm GW}}{\rho_c} h^2
\simeq \frac{h^2}{ \Pi_i \Delta a_i^4}
,
\end{eqnarray}
where $a (T_c) = a_{\rm eq} \times \frac{T_{\rm eq}}{T_c} \sim 10^{-15} \times \frac{100\, {\rm GeV}}{T_c}$ is the scale factor when the thermal bath (SM sector) is at $T_c$, and $\Delta a_i$ are the relative expansion factors of the DM sector against the thermal bath of the SM sector.
For example, to realize the lightest possible DM glueball in the hidden pure Yang-Mills sector [see Eq. (\ref{eq:Gamma_3<H_0})], we need a relative expansion of $\Delta a \sim 10^4$, and this will produce a gravitational wave background with $f \sim 10^{-9}$ Hz and $\Omega_{\rm GW} h^2 \sim 10^{-16}$.
Quoting the result of Ref. \cite{kikuta}, we can probe this HGT with the Square Kilometre Array experiment \cite{ska}, although the estimation is crude.

The third 
is the precision test of the Higgs sector.
The Higgs sector can be studied with high accuracy using 
linear colliders.
This will allow us to examine the possibility of the mini-inflation.
The confirmation of our scenario thus strongly requires the construction of the future international linear collider.

The fourth observable is the CR observation.
As we have investigated, the CRs can probe the unification scale $\Lambda_{\rm GUT}$.
The prospective subject is to measure the CR spectrum above PeV.
We must also determine theoretically the source of the background for the positron excess in the GeV-TeV region.

The final observable expected is the accurate determination of the density profile of the DM halo.
Recent observations suggest that the shape of the DM halo at the innermost of the DM halo has some discrepancy with the result of the simulation of the cold DM with gravitational force only \cite{beyondcollisionlessdm}.
The accurate determination of the halo profile may then provide informations on the self-interaction of the HGT hadrons forming the halo, and find through the hadron level interaction the HGT behind it.


We now summarize this work.
We have studied the DM in the HGT, inspired on the grand unification.
The HGT is natural in the point of view of the unification, but also due to the control of the massive scale by colors, flavors and fermion masses that are less sensitive to the hierarchy problem.
As phenomenological constraints, we have pointed that the stability of the DM halo
 can provide a lower bound on the mass of the DM particle.
The constraint on the DM mass excludes the $SU(2)$ HGT hadrons as the main components of DM when the DM asymmetry is zero.
We have also constrained the unification scale between the SM and the HGT from the CR observation, and have found that this scale is near the currently known GUT scale ($>10^{16}$ GeV).

As a remarkable result, our study suggests the occurrence of the mini-inflation in the SM sector at a high temperature.
The Higgs transition is a good candidate.
This mini-inflation is possible to explain the current baryon-to-photon ratio in the DM pion case.
In the case of pure Yang-Mills HGT or if the HGT hadrons are heavy, an additional mini-inflation above the temperature of leptogenesis is needed.

As future prospects, the confirmation of this scenario may be tried by asking to many experiments.
We are waiting for the detailed analysis of the B-modes,
the stochastic gravitational wave background detections, the high precision test of the Higgs sector using colliders, improvement of the CR detection, and the accurate observation of the density profile at inner regions of the galactic DM halo.

We have thus provided a robust scenario for the thermal evolution of HGT in the point of view of the GUT, and open the way for how to confirm it.

\begin{acknowledgments}
The authors thank 
T. Doi, T. Hatsuda, Y. Ikeda, K. Inayoshi, K. Kohri, S. Nagataki, T. Noumi, N. Ogawa, S. Ohnishi, F. Takahashi and N. Takeda
 for useful discussions and comments.
N. Y. is supported by the RIKEN iTHES Project.
S. F. and S. G. are supported by the Grants-in-Aid for JSPS Fellows (Nos. 26-1329, 24-1458).
\end{acknowledgments}


\begin{thebibliography}{99}

\bibitem{planck}
P. A. R. Ade {\it et al.} (Planck Collaboration), Astron. Astrophys. {\bf 571}, 66 (2014).

\bibitem{zwicky}
F. Zwicky, Helv. Phys. Acta {\bf 6}, 110 (1933).

\bibitem{davis}
M. Davis, G. Efstathiou, C. S. Frenk, and S. D. M. White, Astrophys. J. {\bf 292}, 371 (1985).

\bibitem{clowe}
D. Clowe {\it et al.}, Astrophys. J. {\bf 648}, L109 (2006).

\bibitem{navarro}
J. F. Navarro, C. S. Frenk, and S. D. M. White, Astrophys. J. {\bf 462}, 563 (1996); Astrophys. J. {\bf 490}, 493 (1997).

\bibitem{structureformation}
G. R. Blumenthal, S.M. Faber, J. R. Primack, and M. J. Rees, Nature {\bf 311}, 517 (1984).

\bibitem{macho}
P. Tisserand {\it et al.} (EROS-2 Collaboration), Astron. Astrophys. {\bf 469}, 387 (2007).

\bibitem{dmreview}
G. Bertone, D. Hooper, and J. Silk, Phys. Rep. {\bf 405}, 279 (2005);
J. L. Feng, Ann. Rev. Astron. Astrophys. {\bf 48}, 495 (2010).

\bibitem{susydm}
G. Jungman, M. Kamionkowski, and Kim Griest, Phys. Rept. {\bf 267}, 195 (1996);
N. Arkani-Hamed, S. Dimopoulos, and G. R. Dvali, Phys. Rev. D {\bf 59}, 086004 (1999).

\bibitem{inertdm}
L. Lopez Honorez, E. Nezri, J. F. Oliver, and M. H. G. Tytgat, JCAP {\bf 0702}, 028 (2007).

\bibitem{extradm}
D. Hooper and S. Profumo, Phys. Rep. {\bf 453}, 29 (2007).

\bibitem{axiondm}
J. Preskill, M. B. Wise, and F. Wilczek, Phys. Lett. B {\bf 120}, 127 (1983);
L. F. Abbott, Phys. Lett. B {\bf 120}, 133 (1983).

\bibitem{sterileneutrinodm}
S. Dodelson and L. M. Widrow, Phys. Rev. Lett. {\bf 72}, 17 (1994).

\bibitem{gut}
H. Georgi and S. L. Glashow, Phys. Rev. Lett. {\bf 32}, 438 (1974).

\bibitem{carlson}
E. D. Carlson, M. E. Machacek, and L. J. Hall, Astrophys. J. {\bf 398}, 43 (1992).

\bibitem{secluded}
M. Pospelov and A. Ritz, Phys. Lett. B {\bf 662}, 53 (2008); Phys. Lett. B {\bf 671}, 391 (2009);
Y. Hochberg, E. Kuflik, T. Volansky, and J. G. Wacker, Phys. Rev. Lett. {\bf 113}, 171301 (2014).

\bibitem{hidden}
J. L. Feng, H. Tu, and H.-B. Yu, JCAP 10 (2008) 043;
S. J. Lonsdale and R. R. Volkas, Phys. Rev. D {\bf 90}, 083501 (2014).

\bibitem{compositedm}
W. Shepherd, T. M. P. Tait, and G. Zaharijas, Phys. Rev. D {\bf 79}, 055022 (2009);
D. S. M. Alves, S. R. Behbahani, P. Schuster, and J. G. Wacker, J. High Energy Phys. 1006 (2010) 113;
T. Appelquist {\it et al.} (Lattice Strong Dynamics Collaboration), Phys. Rev. D {\bf 89}, 094508 (2014).

\bibitem{chiral}
S. Bhattacharya, B. Meli\'{c}, and J. Wudka, J. High Energy Phys. 1402 (2014) 115;
J. M. Cline, Z. Liu, G. Moore, and W. Xue, arXiv:1312.3325 [hep-ph].

\bibitem{glueballino}
K. K. Boddy, J. L. Feng, M. Kaplinghat, and T. M. P. Tait, Phys. Rev. D {\bf 89}, 115017 (2014).

\bibitem{lhc}
V. Khachatryan {\it et al.} (CMS Collaboration), arXiv:1408.3583 [hep-ex];
G. Aad {\it et al.} (ATLAS Collaboration), arXiv:1410.4031 [hep-ex].

\bibitem{directdetection}
Z. Ahmed {\it et al.} (CDMS-II Collaboration), Science {\bf 327}, 1619 (2010);
E. Aprile {\it et al.} (XENON100 Collaboration), Phys. Rev. Lett. {\bf 109}, 181301 (2012).

\bibitem{ams-02}
L. Accardo {\it et al.} (AMS Collaboration), Phys. Rev. Lett. {\bf 113}, 121101 (2014).

\bibitem{pamela}
O. Adriani {\it et al.} (PAMELA Collaboration), Nature {\bf 458}, 607 (2009); Phys. Rev. Lett. {\bf 111}, 081102 (2013).

\bibitem{fermilat}
M. Ackermann {\it et al.} (Fermi LAT Collaboration), Phys. Rev. Lett. {\bf 108}, 011103 (2012).

\bibitem{icecube}
M. G. Aartsen {\it et al.} (IceCube Collaboration), Science {\bf 342}, 1242856 (2013).

\bibitem{sakharov}
A. D. Sakharov, Pisma Zh. Eksp. Teor. Fiz. {\bf 5}, 32 (1967) [JETP Lett. {\bf 5}, 24 (1967)].

\bibitem{mccullough}
W. Detmold, M. McCullough, A. Pochinsky, arXiv:1406.2276 [hep-ph]; arXiv:1406.4116 [hep-lat].

\bibitem{strong}
A. W. Strong and I. V. Moskalenko, Astrophys. J. {\bf 509}, 212 (1998).

\bibitem{baltz}
E. A. Baltz and J. Edsj\"{o}, Phys. Rev. D {\bf 59}, 023511 (1998).

\bibitem{ibe}
M. Ibe, S. Iwamoto, S. Matsumoto, T. Moroi, and N. Yokozaki, J. High Energy Phys. 1308 (2013) 029 (arXiv:1304.1483 [hep-ph]).

\bibitem{arkani-hamed}
N. Arkani-Hamed, D. P. Finkbeiner, T. R. Slatyer, and N. Weiner, Phys. Rev. D {\bf 79}, 015014 (2009).

\bibitem{sommerfeld}
A. Sommerfeld, Annalen der Physik, {\bf 403}, 257 (1931).

\bibitem{bec}
C. G. Boehmer and T. Harko, JCAP 0706 (2007) 025.

\bibitem{ngmodes}
H. Watanabe and H. Murayama, Phys. Rev. Lett. {\bf 108}, 251602 (2012);
Y. Hidaka, Phys. Rev. Lett. {\bf 110}, 091601 (2013).

\bibitem{beyondcollisionlessdm}
M. Vogelsberger, J. Zavala, and A. Loeb, Mon. Not. Roy. Astron. Soc. {\bf 423}, 3740 (2012); 
M. Rocha, A. H. G. Peter, J. S. Bullock, M. Kaplinghat, S. Garrison-Kimmel, J. Onorbe, and L. A. Moustakas, Mon. Not. Roy. Astron. Soc. {\bf 430}, 81 (2013);
J. Zavala, M. Vogelsberger, and M. G. Walker, Mon. Not. Roy. Astron. Soc. {\bf 431}, L20 (2013);
A. H. G. Peter, M. Rocha, J. S. Bullock, and M. Kaplinghat, arXiv:1208.3026 [astro-ph.CO];
S. Tulin, H.-B. Yu, and K. M. Zurek, Phys. Rev. D {\bf 87}, 115007 (2013).

\bibitem{inflation}
K. Sato, Mon. Not. Roy. Astron. Soc. {\bf 195}, 467 (1981);
A. H. Guth, Phys. Rev. D {\bf 23}, 347 (1981);
A. D. Linde, Phys. Lett. B {\bf 108}, 389 (1982).

\bibitem{darkenergy}
Adam G. Riess {\it et al.} (Supernova Search Team Collaboration), Astron. J. {\bf 116}, 1009 (1998);
S. Perlmutter {\it et al.} (Supernova Cosmology Project Collaboration), Astrophys. J. {\bf 517}, 565 (1999);
E. J. Copeland, M. Sami, and S. Tsujikawa, Int. J. Mod. Phys. D {\bf 15}, 1753 (2006).

\bibitem{higgs}
G. Aad {\it et al.}, Phys. Lett. B {\bf 716}, 1 (2012);
S. Chatrchyan {\it et al.}, Phys. Lett. B {\bf 716}, 30 (2012);

\bibitem{qcdinflation}
T. Boeckel and J. Schaffner-Bielich, Phys. Rev. Lett. {\bf 105}, 041301 (2010);
Phys. Rev. D {\bf 85}, 103506 (2012).

\bibitem{leptogenesis}
M. Fukugita and T. Yanagida, Phys. Lett. B {\bf 174}, 45 (1986).

\bibitem{sphaleron}
J. Ambjorn and A. Krasnitz, Phys. Lett. B {\bf 362}, 97 (1995).

\bibitem{bicep}
P. A. R. Ade {\it et al.} (BICEP2 Collaboration), Phys. Rev. Lett. {\bf 112}, 241101 (2014); Astrophys. J. {\bf 792}, 62 (2014).

\bibitem{grojean}
C. Grojean and G. Servant, Phys. Rev. D {\bf 75}, 043507 (2007);

\bibitem{kikuta}
Y. Kikuta, K. Kohri, and E. So, arXiv:1405.4166 [hep-ph].

\bibitem{envelope}
A. Kosowsky, M. S. Turner, and R. Watkins, Phys. Rev. D {\bf 45}, 4514 (1992);
A. Kosowsky and M. S. Turner, Phys. Rev. D {\bf 47}, 4372 (1993);
M. Kamionkowski, A. Kosowsky, and M. S. Turner, Phys. Rev. D {\bf 49}, 2837 (1994).

\bibitem{ska}
https://www.skatelescope.org/project/projecttimeline/

\end{thebibliography}
\end{document}